\documentclass[journal]{IEEEtran}
\usepackage{amsmath}
\usepackage{graphicx}
\usepackage{cite}
\usepackage{amsfonts}
\usepackage{booktabs}
\usepackage{multirow}
\usepackage{flushend}
\usepackage{mathrsfs,comment}
\usepackage{hyperref}
\usepackage{cleveref}

\usepackage{longtable, booktabs}
\usepackage{supertabular}
\usepackage{lscape}
\usepackage{eurosym}
\usepackage{verbatim}
\usepackage{xcolor}

\ifCLASSINFOpdf
\else
\fi
\begin{document}

\title{Reconfigurable Intelligent Surfaces for Wireless Communications: Overview of Hardware Designs, Channel Models, and Estimation Techniques}
\author{
%Mengnan Jian, George C. Alexandropoulos, Ertugrul Basar,\\
%\ \ \ \ \ \ Chongwen Huang, Ruiqi Liu, Yuanwei Liu, and Chau Yuen 
Mengnan Jian, George C. Alexandropoulos,~\IEEEmembership{Senior~Member,~IEEE,}  \\Ertugrul Basar,~\IEEEmembership{Senior~Member,~IEEE,} Chongwen Huang,~\IEEEmembership{Member,~IEEE,} Ruiqi Liu,~\IEEEmembership{Member,~IEEE,} \\Yuanwei Liu,~\IEEEmembership{Member,~IEEE,}  and Chau Yuen,~\IEEEmembership{Fellow,~IEEE}
\thanks{M. Jian and R. Liu are with the Wireless Research Institute, ZTE Corporation, Beijing 100029, China and the State Key Laboratory of Mobile Network and Mobile Multimedia Technology, Shenzhen 518055, China (emails: \{jian.mengnan, richie.leo\}@zte.com.cn).}
\thanks{G. C. Alexandropoulos is with the Department of Informatics and Telecommunications,
National and Kapodistrian University of Athens, Athens 15784, Greece (e-mails: alexandg@di.uoa.gr).}
\thanks{E. Basar is with the Communications Research and Innovation Laboratory (CoreLab), Department of Electrical and Electronics Engineering, Ko\c{c} University, Sariyer 34450, Istanbul, Turkey. (e-mail: ebasar@ku.edu.tr)}
\thanks{C. Huang is with College of Information Science and Electronic Engineering, Zhejiang University, Hangzhou 310027, China (e-mail: chongwenhuang@zju.edu.cn).}
\thanks{Y. Liu is with the School of Electronic Engineering
and Computer Science, Queen Mary University of London, London E1 4NS, UK (email: yuanwei.liu@qmul.ac.uk)}
\thanks{C. Yuen is with the Engineering Product Development (EPD) Pillar, Singapore University of Technology and Design, 487372 Singapore (e-mail: yuenchau@sutd.edu.sg).}
}

% The paper headers
%\markboth{Journal of \LaTeX\ Class Files,~Vol.~14, No.~8, August~2015}%
%{Shell \MakeLowercase{\textit{et al.}}: Bare Demo of IEEEtran.cls for IEEE Journals}
% The only time the second header will appear is for the odd numbered pages
% after the title page when using the twoside option.
% 
% *** Note that you probably will NOT want to include the author's ***
% *** name in the headers of peer review papers.                   ***
% You can use \ifCLASSOPTIONpeerreview for conditional compilation here if
% you desire.

% If you want to put a publisher's ID mark on the page you can do it like
% this:
%\IEEEpubid{0000--0000/00\$00.00~\copyright~2015 IEEE}
% Remember, if you use this you must call \IEEEpubidadjcol in the second
% column for its text to clear the IEEEpubid mark.

% use for special paper notices
%\IEEEspecialpapernotice{(Invited Paper)}

% make the title area
\maketitle

\begin{abstract}
The demanding objectives for the future sixth generation (6G) of wireless communication networks have spurred recent research efforts on novel materials and radio-frequency front-end architectures for wireless connectivity, as well as revolutionary communication and computing paradigms. Among the pioneering candidate technologies for 6G belong the reconfigurable intelligent surfaces (RISs), which are artificial planar structures with integrated electronic circuits that can be programmed to manipulate the incoming electromagnetic field in a wide variety of functionalities. Incorporating RISs in wireless networks has been recently advocated as a revolutionary means to transform any wireless signal propagation environment to a dynamically programmable one, intended for various networking objectives, such as coverage extension and capacity boosting, spatiotemporal focusing with benefits in energy efficiency and secrecy, and low electromagnetic field exposure. 

Motivated by the recent increasing interests in the field of RISs and the consequent pioneering concept of the RIS-enabled smart wireless environments, in this paper, we overview and taxonomize the latest advances in RIS hardware architectures as well as the most recent developments in the modeling of RIS unit elements and RIS-empowered wireless signal propagation. We also present a thorough overview of the channel estimation approaches for RIS-empowered communications systems, which constitute a prerequisite step for the optimized incorporation of RISs in future wireless networks. Finally, we discuss the relevance of the RIS technology in the latest wireless communication standards, and highlight the current and future standardization activities for the RIS technology and the consequent RIS-empowered wireless networking approaches.
\end{abstract}

% Note that keywords are not normally used for peerreview papers.

% For peer review papers, you can put extra information on the cover
% page as needed:
% \ifCLASSOPTIONpeerreview
% \begin{center} \bfseries EDICS Category: 3-BBND \end{center}
% \fi
%
% For peerreview papers, this IEEEtran command inserts a page break and
% creates the second title. It will be ignored for other modes.
\IEEEpeerreviewmaketitle

\section{Introduction}
Being on the verge of the final decision on the content of Release $18$ in the $3$rd Generation Partnership Project (3GPP), known also as the first $5$th generation (5G) advanced standard, a number of promising topics have been already discussed \cite{3GPP}, such as the evolution of downlink multiple-input multiple-output (MIMO) and non-terrestrial networks, as well as the inclusion of artificial intelligence (AI) and machine learning in network management \cite{5GPPP_AIML,9148536}, integrated access and backhaul \cite{5G_Americas,9148276}, duplex operation \cite{FD_MIMO}, and uplink/mobility enhancements. Despite the massive roll-out of 5G networks worldwide, which provide higher flexibility and spectrum/energy efficiency compared to their $4$th generation (4G) counterparts, it is inevitable that the $6$th Generation (6G) of wireless networks, expected around $2030$, and beyond will require more radical communication paradigms, especially at the physical layer and multi-access edge computing. 

As per the wide consensus, 5G can be regarded as an evolution of 4G, in terms of physical-layer technologies adopted, while a true revolution can be realized by 6G by enabling Internet-of-everything/Internet-of-sensing applications. According to the recent white paper released by one of the major industrial bodies \cite{Samsung}, some dominant applications of 6G will be truly immersive virtual reality, mobile holography, and digital replica. From the perspective of communication engineering, these science-fiction-type applications will require user experienced and peak data rates of $1$ Gbps and $10$ Gbps, which are $10$- and $50$-fold higher that that of 5G, respectively, along with substantially higher spectrum and energy efficiency. Despite the significant efforts and developments in the physical layer, it is inevitable that we will need to resort more radical communication paradigms to level up these major key performance indicators. Similar to the standardization of 5G which had long past, promising candidate technologies are raising to satisfy the challenging future requirements of 6G wireless networks, including TeraHertz (THz)/visible light communications, AI-empowered communications, wireless power transfer, integrated non-terrestrial networks, cell-free massive MIMO systems, index modulation, advanced waveforms, and smart radio environments enabled/assisted by reconfigurable intelligent surfaces (RISs). 

Communication through RISs appear as a revolutionary paradigm in which the end-to-end wireless propagation channel among end terminals can be controlled by network operators \cite{ChritoLI2018,Huang_2019,Basar_Access_2019,Marco2019,qignqingwu2019,WavePropTCCN,SimRIS3,alexandg_2021,RISE6G_COMMAG,rise6g,Basar_Poor}. Software-controlled RISs are able to manipulate their impinging signals over-the-air to improve the performance of target communication systems, particularly suffering from severe attenuation and blockage. In the past decade, there has been a growing interest in novel communication paradigms that exploit the implicit randomness of the propagation environment. In this context, index modulation that exploits the indices of building blocks of target communication systems to convey information, appeared as an intriguing alternative to well-established MIMO modes and waveforms used in standards \cite{Basar_2017}. RIS-empowered communications take this paradigm one step beyond, by exploiting the wireless channel itself, not only to convey information as in index modulation schemes, but also to boost various communication objectives, such as the error performance, capacity, secrecy, outage, and energy efficiency. In other words, smart radio environments enabled by software-defined RISs can be regarded as one-step beyond software-defined wireless communication networks.

An RIS is a specifically designed man-made surface of electro-magnetic (EM) material that is electronically controlled with integrated electronics and has unnatural wireless communication capabilities. In the most commonly considered case, the large number of small-sized, low-cost, and almost passive elements that comprise an RIS can simply modify the incident signal over-the-air to improve the signal coverage and/or quality \cite{huang2020holographic}. This technology is different from existing massive MIMO transceivers \cite{mMIMO} and active beamformers \cite{Molisch_HBF_2017_all}, as well as the cooperative relaying \cite{Cooperative,HoVan_relay_selection,comparative_study} and backscatter communication \cite{Backscatter} paradigms. The most distinguishable property of RISs is their cheap and nearly passive panel of unit cells, each being capable of applying a controllable phase shift to the impinging field (e.g., via two-state reflection coefficients in one of the first RIS designs in \cite{Kaina_metasurfaces_2014}). When used as pure tunable reflectors, RISs do not require radio frequency (RF) signal processing, amplification, filtering, mixing, or up/down-conversion. In light of these, RIS-empowered communication can be regarded as a alternative/complementary technology to massive MIMO and active relaying systems.

Remarkable developments in the field of RIS-empowered communications has been witnessed in the past 3 years, and the underlying RIS technology has become mature enough for both academia and industry to explore its promising future use-cases. Particularly, within the context of RIS wireless systems, researchers have focused on joint active and passive beamforming problems \cite{Huang_2019,Wu_2019,Di_2020}, channel modeling under different conditions \cite{bjornson_Power_2020,Tang_2021,Basar_2021,Danufane_2021,PhysFad} and performance analyses \cite{Nadeem2020,Selimis2021,Moustakas_RIS}, physical layer security \cite{Cui_2019,Chen_2019,PLS_Kostas}, non-orthogonal multiple access schemes (NOMA) \cite{Ding_2019,Hou_2020,Nahhal_2021}, non-coherent modulation \cite{mine_NC_2021,Kun_NC_2022}, vehicular networks \cite{Dampahalage_2020,Doppler}, hybrid passive and sensing designs \cite{HRIS,HRIS_SPAWC,HRIS_Mag}, active and joint reflective-transmissive designs \cite{Active_Passive_2021,Star_2021}, deep learning solutions \cite{huang2019spawc,Samarakoon_2020,Taha_2021,Wang2021jointly_all,DeepRIS_2022}, reflection modulation schemes \cite{Guo_2020,Lin_2021,Yuan_2021}, real-word experiments with prototyping \cite{Dai_2020, Tang_2021, Jin_2021}, along with many other interesting application areas.

As in any other communication technology, channel modeling is an integral part of the RIS research to understand the most convincing use cases of RISs in future wireless networks. Within this perspective, one has to not only consider the physical propagation effects in different environments, such as different path loss and scattering behavior of large indoor office or street canyon environments, but also the effect of the RIS on the end-to-end channel model, such as near-field effects for RISs in close proximity and potential spatial correlation \cite{NakagamiTWC_2009,alexandg_MRT} as well as mutual coupling \cite{Nossek,alexandg_ESPARs} among RIS elements. The architecture of the RIS, such as reflect-array-based or metasurface-based, plays an important role to the underlying channel models. Moreover, modified line-of-sight (LOS) probabilities as well as path loss exponents can be derived considering potential locations of RISs in practical wireless environments.

When considering smart wireless environments with almost passive RISs, the knowledge of channel states or how to perform channel estimation in highly dynamic environments are quite challenging. Unfortunately, nearly passive reflective RIS architectures like \cite{Subrt_2012,Kaina_metasurfaces_2014} cannot perform any form of parameter estimation at their side, and the problem becomes the estimation of the cascaded channel between the transmitter and the receiver through the RIS. Furthermore, the increasing size of the RIS increases the channel estimation burden and dedicated channel-bearing signaling might be required between the network controller and the RIS to adjust its configuration in real-time. At this point, standalone RIS architectures equipped with AI modules might provide promising solutions to configure the RIS on the go along with efficient channel estimation protocols, which have low signaling overhead for RIS-empowered systems (e.g., \cite{low_overhead}).

The last years, there has been increasing interest in transceiver hardware architectures (see, e.g., \cite{WavePropTCCN, Molisch_HBF_2017_all} and references therein), and researchers have explored effective schemes to provide reconfigurability, such as analog beamforming networks, large arrays of inexpensive antennas equipped with external phase shifters, reconfigurable loads with varactor diodes, or even reconfigurable reflecting elements with PIN diodes. Even more recently \cite{Amplifying_RIS}, active RIS systems including a single power amplifier have been proposed for offering reflection amplification. Their aim is to challenge the double path loss effect of passive RIS systems, and promising results were reported in terms of both achievable rate performance and energy efficiency. Considering the higher cost, complexity, and power consumption of these active RIS architectures, interesting new problems arise in terms of system optimization and hybrid RIS system design. Furthermore, practical analyses on mutual coupling as well as the correlation between reflection amplitudes and phases are gaining momentum.

In line with the latest rich research and development, as well as the ongoing booming interests in RIS-empowered wireless networking, this survey article presents various up-to-date advances in the RIS technology, focusing particularly on the taxonomy of the available RIS hardware architectures and their operation modes, the diverse channel modeling approaches including multi-user NOMA systems, as well as the plethora of the channel acquisition techniques. In particular, in Section~\ref{sec:RIS_Hardware}, we present a taxonomy of the available RIS hardware designs and their available implementations, while Section~\ref{sec:Models} includes the overview of the up-to-date channel models for wireless communications empowered by RISs. Section~\ref{sec:Channel_Estimation} discusses the available channel estimation approaches in RIS-empowered wireless communication systems, while the relevance of the RIS technology in the latest standards together the topic's road ahead are discussed in Section~\ref{sec:Standardization}. The concluding remarks of the paper are included in Section~\ref{sec:Conclusions}.

\begin{figure*}[ht] \vspace{-0.0cm} \hspace{-0.0cm}
            \centering
            \includegraphics[width=17.9cm]{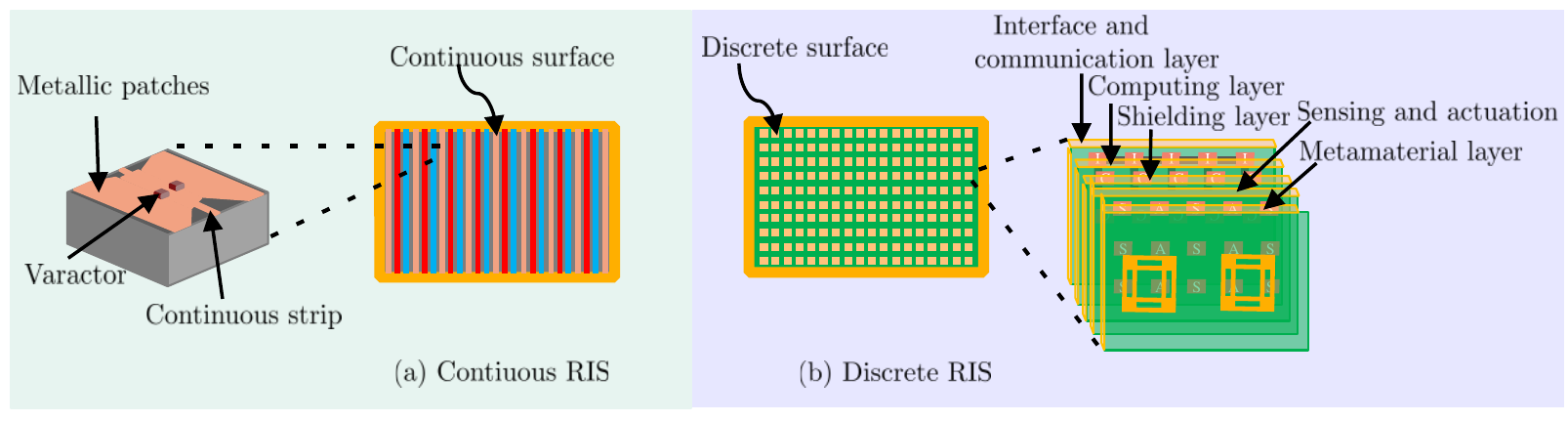} \vspace{-0.0cm}
            \caption{A schematic view of RIS hardware structures \cite{huang2020holographic}, including (a) continuous and (b) discrete implementations .}
\end{figure*}

\section{RIS Hardware Architectures}\label{sec:RIS_Hardware}
In this section, we present the up-to-date RIS hardware architectures and some indicative fabrication methodologies. Compared with the currently used transceiver technologies in wireless networks, the most distinctive characteristics of the RIS concept lie in making the environment controllable by providing the possibility of fully shaping and controlling the EM response of the environmental objects that are distributed throughout the network. An RIS structure is usually intended to operate as a signal source or wave collector with reconfigurable characteristics, especially for application scenarios where it is used as a passive reflector with the objective of improving the communication performance. Existing RISs can be roughly categorized as contiguous and discrete RISs, while on the basis of power consumption, RIS also can be classified into the active and passive types. We also discuss the different operation modes that an RIS can be used, identifying the hardware requirements per distinctive mode.

\subsection{Hardware Designs}
\subsubsection{Active RISs} The term active RIS is adopted when energy-intensive RF circuits and consecutive signal processing units are embedded in the surface \cite{holobeamforming,Marzetta2019}. On another note, active RIS systems comprise a natural evolution of conventional massive MIMO systems, by packing more and more software-controlled antenna elements onto a two-dimensional (2D) surface of finite size \cite{9324910}. In \cite{husha_LIS1}, the authors name an active RIS as a large intelligent surface (LIS), where the element spacing reduces when the element number increases. A compact integration of a large number of tiny active RIS elements with reconfigurable processing networks can realize a practical continuous antenna aperture. The active RIS structure can be used to transmit and receive signals across the entire surface utilizing the hologram principle \cite{holobeamforming,Marzetta2019}. The discrete photonic antenna array is another practical implementation of active RISs. It integrates active optical-electrical detectors, converters, and modulators for performing transmission, reception, and conversion of optical or RF signals \cite{holobeamforming}. In \cite{hardware2020icassp}, RISs consisting of passive RIS elements and a single receive RF chain for baseband measurements were presented. Very recently in \cite{dong2021active}, an active RIS structure where each reflecting element is equipped with a power amplifier is proposed.  

\subsubsection{Passive RISs}
Passive RIS acts like a passive metal mirror or wave collector which can be programmed to change an impinging EM field in a customizable way\cite{Huang_2019,Marco2019}. Compared with its active counterpart, a passive RIS is usually composed of low-cost and almost passive elements that do not require dedicated power sources. Their circuitry and embedded sensors can be powered with energy harvesting modules, an approach that has the potential of making them truly energy neutral. Regardless of their specific implementations, what makes the passive RIS technology attractive from an energy consumption standpoint, is their capability to shape radio waves impinging upon them, forwarding the incoming signal without employing any power amplifier nor RF chain, and even without applying sophisticated signal processing. Moreover, passive RISs can work in full duplex mode without significant self interference or increased noise level, and require only low-rate control link or backhaul connections. Finally, passive RIS structures can be easily integrated into the wireless communication environment, since their extremely low power consumption and hardware costs allow them to be deployed into building facades, room and factory ceilings, laptop cases, or even human clothing \cite{Huang_2019,Marco2019}.

\subsubsection{Discrete RISs}
 A discrete holographic multiple input multiple output surface (HMIMOS) is usually composed of many discrete unit cells made of low-power and software-tunable metamaterials. The means to electronically modify the EM properties of the unit cells range from off the shelves electronic components to using liquid crystals, microelectromechanical systems or even electromechanical switches, and other reconfigurable metamaterials. This structure is substantially different from the conventional MIMO antenna array. One embodiment of a discrete surface is based on discrete `meta-atoms' with electronically steerable reflection properties \cite{ChritoLI2018}. As mentioned earlier, another type of discrete surface is the active one based on photonic antenna arrays. 
 
 \subsubsection{Contiguous RISs}
Integrating a virtually infinite number of elements into a limited surface area, a contiguous RIS can thus form a spatially continuous transceiver aperture \cite{holobeamforming,Marzetta2019}. Compared with discrete RISs, contiguous RISs have some essential differences from the perspectives of implementation and hardware, as will be described in the sequel in the fabrication methodologies subsection. 

\subsection{Operation Modes}

\subsubsection{Reflecting RISs}
The concept of the RIS-empowered smart wireless environments \cite{ChritoLI2018,Marco2019,Huang_2019} initially considered only passive RISs with almost zero power consumption unit elements \cite{WavePropTCCN}. Their envisioned prominent role lies on the capability of the surface to reconfigure the reflection characteristics of its elements, enabling programmable manipulation of incoming EM waves in a wide variety of functionalities. It is  essential to achieve a fine-grained control over the reflected EM field for quasi-free space beam manipulation so as to realize accurate beamforming. 
Meta-atoms of sub-wavelength size are a favorable choice, although inevitable strong mutual coupling, and well-defined gray-scale-tunable EM properties exist.
%This fact motivated researchers to rely on meta-atoms of sub-wavelength size, despite inevitable strong mutual coupling [18] among meta-atoms, and well-defined gray-scale-tunable EM properties.

Conversely, in rich scattering environments \cite{alexandg_2021}, the wave energy is statistically equally spread throughout the wireless medium. The ensuing ray chaos implies that rays impact the RIS from all possible, rather than one well-defined, directions. The goal becomes the manipulation of as many ray paths as possible, which is different from the common goal of creating a directive beam. This manipulation has two kind of aims, including tailoring those rays to create constructive interference at a target location and stirring the field efficiently. These manipulations can be efficiently realized with RISs equipped with half-wavelength-sized meta-atoms, enabling the control of more rays with a fixed amount of electronic components (PIN diodes, etc.).

\begin{figure}[!t]
	\begin{center}
		\includegraphics[width=\linewidth]{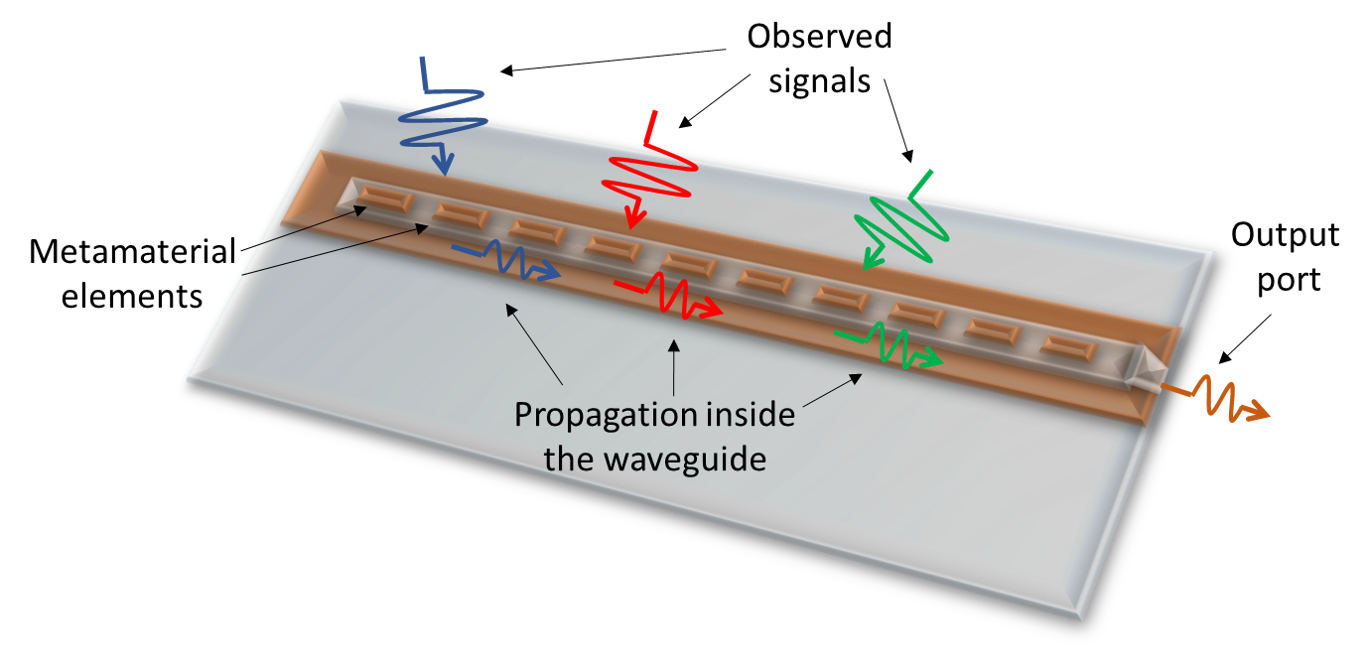}
		\caption{Schematic illustration of a microstrip for the DMA design in \cite{9324910}. The microstrip's output port is connected to a receive RF chain for baseband processing. The port can be alternatively fed by the output of a transmit RF chain to realize analog transmit beamforming.}
		\label{DMA}
	\end{center}
\end{figure}
\subsubsection{Metasurface-Based Antennas}
As an efficient realization of extreme massive antenna arrays, dynamic metasurface antennas (DMAs) have been recently proposed \cite{9324910} to pave the way. DMAs have beam tailoring capabilities and facilitate processing of the transmitted and received signals in the analog domain. DMAs work in a dynamically configurable manner with simplified transceiver hardware. Additionally, compared with conventional antenna arrays, DMA-based architectures require much less power and cost. In this way, eliminating the need for complicated corporate feed and / or active phase shifters become possible. Another promising advantage of DMAs is that they can comprise massive numbers of tunable metamaterial-based antenna elements fitting into small physical areas and providing wide range of operating frequencies.

In \cite{9324910}, a DMA architecture that consists of multiple separate waveguide-fed element arrays with each connected to a single input/output port is discussed. A large number of radiating elements can be accommodated in waveguides, and the sub-wavelength spaced character allows each input/output port to feed a multitude of possibly coupled radiators. For 2D waveguides, a scattered wave from each element propagates in all directions. Since the proposed waveguide is typically designed to be single mode and the wave can only propagate along one line, its analysis is much easier than 2D waveguides. Furthermore, ensuring isolation between different ports is easier in 1D waveguides than in multiple ports of a 2D waveguide. A common implementation of 1D waveguides is based on microstrips, as depicted in Fig.~\ref{DMA}.

\subsubsection{Receiving RISs}
The DMA principle was also conceptualized in the parallel work \cite{hardware2020icassp} for enabling RISs with baseband reception capability in order to perform explicit or implicit channel estimation with minimal number of receive RF chains. In fact, \cite{RX_RIS_Localization} presented a signal direction estimation approach with a single RF chain, while \cite{hardware2020icassp} designed a matrix-completion-based channel estimation technique with relatively short training requirements, requiring much fewer receive RF chains than RIS elements.

\begin{figure}[!t]
	\begin{center}
		\includegraphics[width=0.95\linewidth]{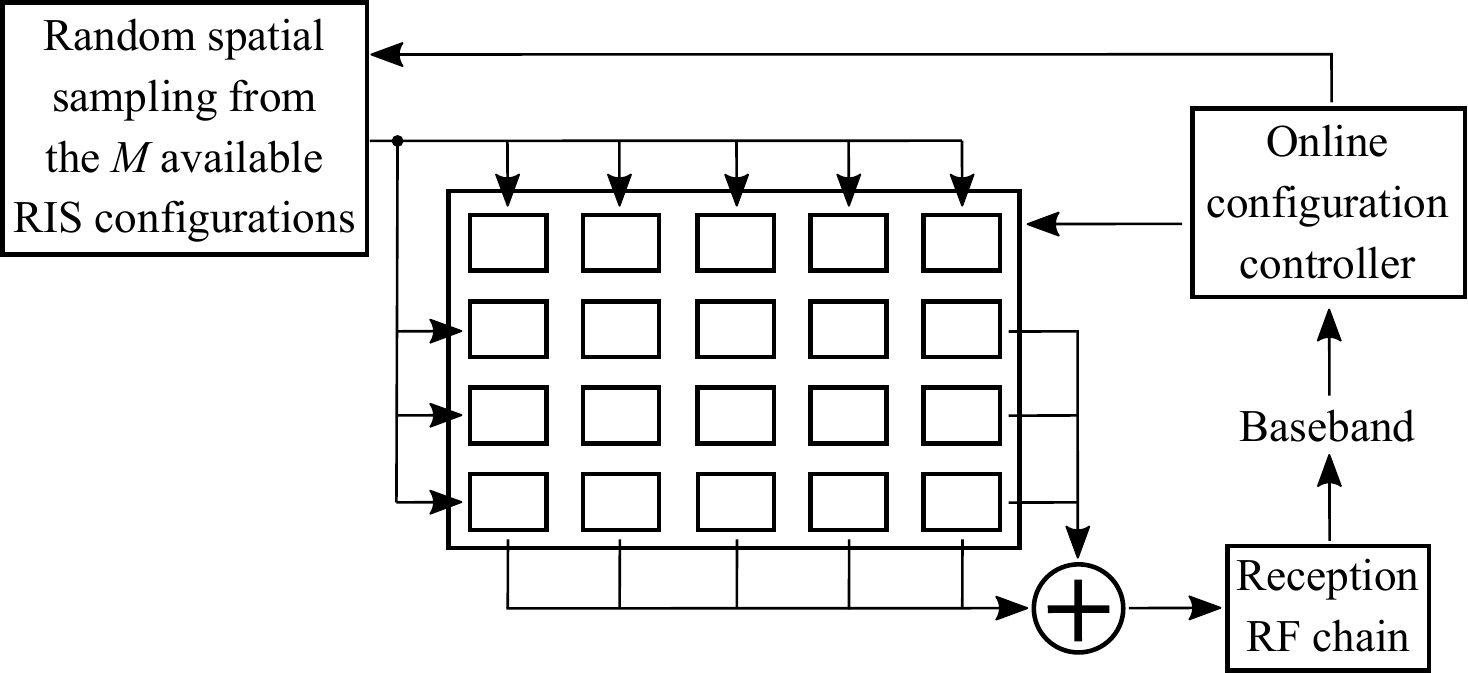}
		\caption{Block diagram of the receiving RIS hardware architecture in \cite{hardware2020icassp} including a single active reception RF chain for explicit channel estimation at the RIS side.}
		\label{RX_RISs}
	\end{center}
\end{figure}

The receiving RIS hardware architecture, which is illustrated in Fig.~\ref{RX_RISs}, connects the outputs of the RIS elements to a single reception RF chain. The hardware includes a low noise amplifier, a mixer downconverting the signal from RF to baseband, and an analog-to-digital converter. Each impinging EM training signal at the meta-elements is received in the RF domain with one of the $M$ available RIS configurations included in the random sampling unit. This configuration is selected by random spatial sampling, which is essential for efficient channel reconstruction \cite{Vlachos2019}. 

\subsubsection{Simultaneously Reflecting and Sensing RISs}
A hybrid meta-atom, which can simultaneously reflect a portion of the impinging signal in a programmable way while another portion of it can be fed to a sensing unit, has been recently designed in \cite{HRIS}. The so-called hybrid RISs are realized by adding a waveguide to couple to each meta-atom. Each waveguide can be connected to an RF chain. This makes it possible to locally process a portion of the received signals in the digital domain. However, the coupling of the RIS elements to the waveguides makes it impossible for the incident wave to be perfectly reflected. Actually, the coupling level determines the ratio of the reflected energy to the absorbed energy. Its footprint can be reduced and the coupling to the sampling waveguide can be mitigated by keeping this waveguide near cutoff. Such hybrid RISs were recently considered in \cite{HRIS_Mag,HRIS_SPAWC} for facilitating explicit and implicit channel estimation as well as flexible network management with reduced overhead compared to cascaded channel estimation with passive RISs. 
\begin{figure}[!t]
	\begin{center}
		\includegraphics[width=\linewidth]{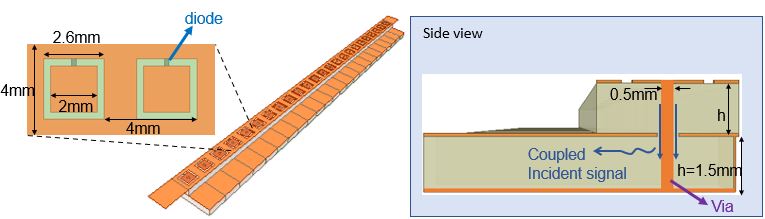}
		\caption{Portion of the hybrid RIS design in \cite{HRIS} with integrated sensing capability.}
		\label{HRIS}
	\end{center}
\end{figure}

In Fig.~\ref{HRIS}, the designed hybrid RIS (HRIS) of \cite{HRIS} is schematically illustrated. It consists of mushroom structures, each loaded with a varactor diode; this reconfigurable capacitance results in a reconfigurable resonance frequency, and consequently, a reconfigurable effective impedance. This element was selected since it provides a simple mechanism to realize high reflectivity with reconfigurable phases.To address each meta-atom independently, as required for forming desired refection patterns, the via of the mushroom structure extends
through the bottom conductive plate. An annular slot separates the via from the
ground plane beneath the substrate. This annular slot allows for coupling the incident wave to another layer. In previous designs, an RF choke (in the form of a radial stub) was placed at the end of this via to diminish RF coupling. In \cite{HRIS}, however, this annular slot was exploited to sense the incident wave. This sensing information can be then used for various network management tasks.

 \subsubsection{Simultaneously Transmitting and Reflecting RISs}
\begin{figure}[!t]
	\begin{center}
		\scalebox{0.5}{\includegraphics{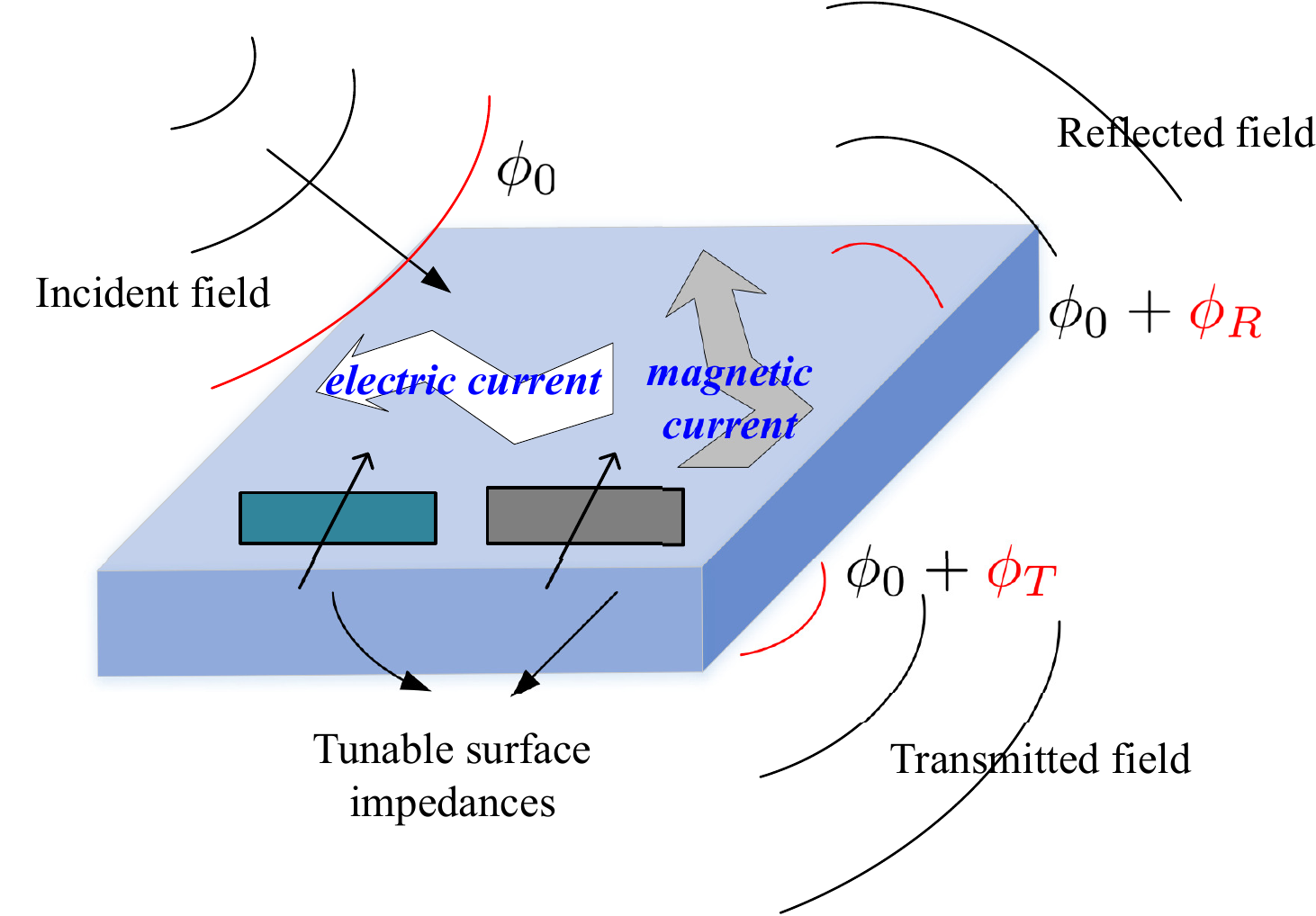}}
		\caption{Schematic illustration of the STAR-RIS element, where $\phi_0$, $\phi_R$, and $\phi_T$ denote the phases of the incident, reflected, and transmitted signals, respectively.}
		\label{star}
	\end{center}
\end{figure}

The concept of simultaneously transmitting and reflecting RISs (STAR-RISs) was recently presented in~\cite{xu_star}. This new type of RIS variant is also referred to as the intelligent omni-surface (IOS)~\cite{IOS_zhang}, allowing wireless signals incident on the surface to be simultaneously reflected and transmitted. STAR-RISs can assist in achieving a full-space reconfigurable wireless environment that has a $360$-degree coverage~\cite{liu_star}. In Fig.~\ref{star}, the schematic illustration of the STAR-RIS element is presented, where the tunable surface impedance can be configured to adjust the intensities and distributions of the surface electric and magnetic currents. This, in principle, enables an independent control of both the reflecting phase $\phi_R$ and the tranmsission phase $\phi_T$ appearing at a STAR-RIS.

%(STAR-RISs hardware prototypes)
The key to achieving tunability for both the reflected and transmitted (refracted) signals is that each element has to support both electric and magnetic currents. As reported in~\cite{xu_vtmag}, several prototypes have been implemented which support simultaneous reflection and transmission, e.g., the transparent dynamic metasurface designed by researchers at NTT DOCOMO~\cite{DOCOMO} and the PIN diode empowered by the STAR-IOS prototype~\cite{chao_star,IOS}. Similar to the approach of studying conventional reflecting RISs, the hardware modeling of STAR-RISs comes with additional challenges, because transmitted (refracted) signals need to be considered. The authors of~\cite{xu_star} studied a general hardware model and near/far-field channel models for STAR-RISs, where the reflection and transmission coefficients of the unit elements were assumed to be independently adjustable. However, in practical cases, the unit elements at STAR-RISs have to be passive-lossless to maintain good scalability. Therefore, the authors of \cite{xu_correlated,liu_coupled} considered the constraints that should be imposed on the reflection and transmission coefficients of a given passive STAR-RIS element.

\subsubsection{Amplifying RISs}
An amplifying RIS design that incorporates a single power amplifier (PA), enabling reflection amplification, was recently proposed in \cite{Amplifying_RIS}. Its motivation lies on the double path loss problem and the lossy reflection with passive reflective RISs. In fact, only when the passive RISs are positioned close to terminals can they perform efficiently. Otherwise, they cannot go beyond being a supportive technology for future wireless communications systems. According to this very recent amplifying RIS design, the impinging signal is received by a portion of the RIS. Then, the received EM field is phase configured and fed to the single power amplifier, who in turns feeds it to the remaining portion of the RIS that reflects it with the controllable phase configuration. The amplifying design works solely in the RF domain similar to the waveguide-based approach in \cite{HRIS}. This is different from the full-duplex multi-antenna decode-and-forward relays, as down-conversion and baseband processing is performed \cite{HybridRIS}.

\subsection{Fabrication Methodologies}
There are various fabrication techniques for RISs including electron beam lithography at optical frequencies, focused-ion beam milling, interference and nanoimprint lithography, as well as direct laser writing or printed circuit board processes at microwaves. Usually, these fabrication techniques will be ascribed to produce two typical apertures, continuous or discrete apertures. In approximately continuous microwave aperture, the meta-particle structure uses the varactor loading technique to broaden its frequency response range, and achieves controllable reflection phase. It is a continuous monolayer metallic structure, and comprises a large number of meta-particles. Each meta-particle contains two metallic trapezoid patches, a central continuous strip, and varactor diodes. By independently and continuously controlling the bias voltage of the varactors, the surface impedance of continuous RISs can be dynamically programmed, and thus manipulate the reflection phase, amplitude states, and the phase distribution over a wide range of frequency bands \cite{Akyildiz2018mag}. It should be highlighted that this impedance pattern is a map of the hologram, and can be calculated directly from the field distribution of the provided reference wave and reflected object wave. 

In contrast to continuous apertures, another instance of RISs is a realization based on discrete apertures that are usually realized with software-defined metasurface antennas. A general logical structure was proposed in \cite{ChritoLI2018}. Its general unit cell structure contains a metamaterial layer, sensing and actuation layers, shielding layer, computing layer, as well as an interface and communications layer with different objectives. Specifically, the meta-material layer is implemented by graphene materials for delivering the desired EM behavior through a reconfigurable pattern, while the objective of sensing and actuation layer is to modify the behavior of the meta-material layer. The shielding layer is made of a simple metallic layer for decoupling the EM behavior of the top and bottom layers to avoid mutual interference. The computing layer is used to execute external commands from the interface layer or sensors. Finally, the interface and communications layer aim at coordinating the actions of the computing layer and updating other external wireless entities via the reconfigurable interface.

While the development of RISs is in its infancy, basic prototyping work on different kinds of this technology is available already. Various discrete RISs have been developed by the start-up company ``Greenerwave,'' which shows the basic feasibility and effectiveness of the RIS concept using discrete metasurface antennas. In contrast, another start-up company named as ``Pivotalcommware,'' with the investment of Bill Gates, is developing initial commercial products of a contiguous RIS based on low-cost contiguous metasurfaces; this further verifies the feasibility of the RIS concept as well as the advancement of holographic technologies. The ZTE Corporation, a network infrastructure vendor, also demonstrated recently a series of RIS panels which are dual-polarized and capable of $2$-bit phase control, empowered by PIN diodes as well as liquid crystals \cite{ZTE_RIS_demo}. Continuing prototyping development is highly desired to prove the RIS concept with brand new holographic beamforming technologies, potentially unveiling new technical issues which will require further research efforts. In addition, one of the main goals of the ongoing European Union H2020 RISE-6G project \cite{RISE_6G_Online}, which is the first collaborative project planning to design, prototype, and trial radical technological advances based on RISs to forge a new generation of dynamically programmable wireless propagation environments \cite{RISE6G_COMMAG,rise6g}, is to provide low-power and cost-effective RIS hardware designs operating from below $6$ GHz to millimeter wave (mmWave) and sub-THz frequencies.

\section{RIS-Empowered Signal Propagation Modeling}\label{sec:Models}
The accurate modeling of wireless channels is one of the most critical factors enabling wireless technologies, and serves as a core means for validating them before and in conjunction with field trials. As such, it has attracted great interest in the scientific community of wireless communications the recent years. From the viewpoint of characterization, many measurement campaigns \cite{RMS15, METIS} have studied wireless channels from a static point of view. From these
measurement campaigns, we learned that there are some propagation paths available when the LOS component (or another strong path in case of non-LOS propagation) is blocked. However, in real situations such as in street canyon environments, all propagation paths are within the same beam and are blocked simultaneously, leading to a significant performance decrease \cite{WPK14}. The increasing demand for higher frequency bands has recently stimulated the proposal of new models for above 100 GHz \cite{RXK19}, providing first results on angular characterization of the sub-THz channel \cite{PDE19}. On the scenario side, going beyond the classical enhanced mobile broadband, the industrial environment motivated new studies also addressed by 3GPP. Very recently, there exists increasing research interest in propagation channel models that incorporate the change of the wireless environment itself via the inclusion of RISs. In the following, we overview the key latest advances in channel models for wireless communications empowered by RISs.

Dating back to $1992$ to the concept of time reversal \cite{Fink_TR_1992} in conjunction with its very recent feasibility demonstrations in high frequency bands with large available transmission bandwidths \cite{Alexandropoulos_ICASSP,TR_CSCN_2021,TR_WCNC_2022}, as well as the recent interests for RIS-empowered smart wireless environments \cite{ChritoLI2018,Huang_2019,Basar_Access_2019,Marco2019,qignqingwu2019,WavePropTCCN,SimRIS3,alexandg_2021,RISE6G_COMMAG,rise6g,Basar_Poor}, it becomes apparent that controlling waves in complex media is a major topic of interest. In \cite{del2016spatiotemporal}, the authors experimented with an RIS inside a cavity, emulating a rich scattering environment, and showcased that by maximizing the Green's function between two antennas at a chosen time, yields diffraction-limited spatiotemporal focusing. In addition, by changing the photons' dwell time inside the designed cavity, the relative distribution of the spatial and temporal degrees of freedom were modified. Such RIS-enabled tunable rich scattering conditions were recently considered in \cite{alexandg_2021} for analog multipath shaping. In particular, the RIS was deployed to create pulse-like channel impulse responses that can maximize the channel capacity. In \cite{DeepRIS_2022}, the authors used a coupled-dipole-based simulator, which faithfully models the underlying wave
physics in RIS-empowered communications inside cavities, to simulate fast fading conditions in rich scattering environments.  

A stochastic cascaded model for RIS-empowered wireless communications was initially presented in \cite{Huang_2019,qignqingwu2019}, where the impact of the RIS in the signal propagation was encapsulated through the reflection coefficients of its state-tunable unit elements. In this generic model which has been extensively adopted up to date, typical channel models can be incorporated (e.g., Rayleigh, Ricean, and Nakagami-$m$) and the RIS is mainly assumed to be capable of controlling the phases of its unit-amplitude reflection coefficients. Considering realistic circuitry for resonant RIS elements and capitalizing on the transmission line model, \cite{Abeywickrama_2020} presented a practical model for its RIS-induced reflection coefficient, where the phase configuration depends on the amplitude. Very recently in \cite{Katsanos_Lorentzian}, the authors considered a transfer function approach and introduced a physics-compliant model for the Lorentzian frequency response of the RIS elements, which captures the parameters one can externally
control to modify the surface's reflection profile. Based on this RIS frequency response model, an optimization framework for setting the RIS controllable parameters in order to maximize the achievable average sum-rate in the uplink of wideband RIS-empowered multi-user
MIMO systems with orthogonal frequency division multiplexing (OFDM) signaling was presented. A multipath signal propagation channel model based on the cascaded channel model was considered in\cite{RIS_equalization_icassp2021}, where the authors presented an optimization framework for RIS-enabled equalization, aiming to reduce the inter-symbol interference (ISI). 

A two-path propagation model was proposed in \cite{multipath_Mitigation_RIS} for RIS-assisted wireless networks by considering both the direct path from the transmitter to the receiver and the assisted path established by the RIS. The proposed propagation model unveiled that the phase shifts of RISs can be optimized by appropriate configuration for multipath fading mitigation. In particular, four types of RISs with different configuration capabilities were introduced and their performances on improving received signal power in virtue of the assisted path to resist fast fading were compared through extensive simulation results. An end-to-end channel model including frequency selectivity for RIS-enabled wireless communications was presented in \cite{over_the_air_RIS}, together with a framework for the potential of RISs for performing equalization. It has been theoretically shown that the RIS can be adjusted according to the incoming signals to maximize the magnitude of the first channel tap tap, making the ISI terms negligible, hence, maximizing the achievable rate performance. In a similar manner, RISs have been used in \cite{Doppler} to confront the Doppler effect in mobile RIS-empowered communications.

By partitioning the RIS units into several subsets, referred to as tiles, a physics-based model is presented in \cite{najafi_Physics_2020} that also incorporates polarization. The impact of each tile on the wireless channel is modeled using concepts from the radar literature, according to which it is an anomalous reflector. In this case, for a given phase shift, the tile's impact can be derived by solving the corresponding integral equations for the electric and magnetic vector fields. Using this model, each RIS tile is then optimized in two stages. In the offline stage, the RIS units of each tile are jointly designed to support different transmission modes, with every transmission mode effectively corresponding to a given configuration of the phase shifts that the units of the tile apply to an incident EM wave. In the online optimization stage, the best transmission mode of each tile is selected such that a desired quality of service (QoS) criterion is maximized. 

A channel gain expression for planar arrays of arbitrary sizes, taking the varying
distances to the antenna elements, polarization mismatches, and the effective areas into account, was presented in \cite{bjornson_Power_2020} to investigate the spectral efficiency behavior and asymptotic power scaling laws in RIS-empowered communications. It was proved for this deterministic propagation model (i.e., free-space LOS) that an RIS cannot achieve
a higher signal-to-noise ratio (SNR) than MIMO setups when the array sizes are equal, despite the fact that the SNR in the RIS setup grows with the square of its number of elements in the far-field. A closed-form expression showcasing how large an RIS must be to beat conventional massive MIMO or half-duplex massive MIMO relaying was derived. Using physical optics techniques in \cite{ozdogan_Intelligent_2020}, a far-field pathloss model was presented that intends to explain why an RIS, consisting of many elements that individually act as diffuse scatterers, can beamform the signal in a desired direction with a certain beamwidth. 

In \cite{ajam_Modeling_2021}, an analytical channel model for point-to-point RIS-assisted free-space optical (FSO) systems was developed, which is based on the Huygens-Fresnel principle for the intermediate and the far-field. The model determines the reflected electric field and captures the impact of the size, position, and orientation of the RIS, as well as its phase shift profile on the end-to-end channel. The accuracy of the proposed analytical model was validated via simulations and it was shown that, in contrast to models based on the far-field approximation, it is valid even for intermediate distances, which are relevant in practice. The model was exploited to devise RIS sharing protocols in multi-link FSO systems, whose performance was analytically investigated.

Considering an RIS modeled as a sheet of EM material of negligible thickness, and leveraging the general scalar theory of diffraction as well as the Huygens-Fresnel principle in \cite{direnzo_Analytical_2020}, closed-form expressions for the power reflected from an RIS are presented as functions of the size of the RIS, the distance between the transmitter/receiver and the RIS, and the phase shift matrix configured by the RIS. With the aid of the stationary phase method, the authors identify sufficient conditions under which an RIS acts as an anomalous mirror, indicating that the received power decays as a function of the reciprocal of the sum of the distances between the transmitter/receiver and the RIS. It is shown that in short distances, an RIS acts as an anomalous mirror, and in the far distance regime acts as a scatterer. Based on the vector generalization of the Green's theorem, a free-space pathloss model for homogenized RISs, made of sub-wavelength scattering elements and being capable to operate either in reflection or transmission mode, is introduced in \cite{Danufane_2021}. The model is formulated in terms of a computable integral that depends on the transmission distances, the polarization of the radio waves, the size of the RIS, and the desired surface transformation. Closed-form expressions are obtained in two asymptotic regimes that are representative of far-field and near-field deployments. A summary of the available path loss models was recently provided in \cite{ellingson_PathLoss_2021} with some of them being experimentally validated inside an anechoic chamber in \cite{Tang_2021}. Very recently, pathloss models considering THz frequencies are presented \cite{boulogeorgos_Pathloss_2021, dovelos_Intelligent_2021}. In \cite{dovelos_Intelligent_2021}, a near-field channel model is proposed that accounts for the size of the RIS elements in the path loss calculation, as well as in the spherical wavefront of the radiated waves. It is shown that a typical THz RIS is likely to operate in the Fresnel zone, where conventional beamforming is suboptimal and hence can reduce the power gain. 

Capitalizing on the circuit theory of communications \cite{Nossek} and its applications in tunable impedance systems (e.g., in beamforming with single-RF electronically steerable parasitic array radiators \cite{alexandg_ESPARs}), an end-to-end mutual-coupling-aware channel model for free-space signal propagation empowered by an RIS was presented in \cite{gradoni_EndtoEnd_2020}. The model is based on the mutual impedances between all existing radiating elements, i.e., the transmit/receive antennas and the passive scatterers which are all modeled as dipoles. It yields an one-to-one mapping between the voltages
fed into the ports of a transmitter and the voltages measured at
the ports of a receiver, while accounting for the generation and propagation of the EM fields, the mutual coupling among the sub-wavelength unit cells of the RIS, and the interplay between the amplitude and phase response of the unit cells of the RIS.

Very recently in \cite{PhysFad}, the authors introduced a physics-based end-to-end model of RIS-parametrized wireless channels with adjustable fading, which is based on a first-principles coupled-dipole formalism. The proposed model, whose MATLAB source code has been made available online, incorporates the notions of space and causality, dispersion (i.e., frequency selectivity) and the intertwinement of each RIS element’s
phase and amplitude response, as well as any arising mutual coupling effects including long-range mesoscopic correlations. This model has been used for simulating Ricean fading channel and was considered for a prototypical problem of RIS-enabled over-the-air channel equalization in rich-scattering wireless communications. 

In \cite{Basar_2021}, the authors capitalize on the 5G mmWave channel model with random
number of clusters/scatterers \cite{5G_mmWave} and extend it to include an RIS. The proposed model is valid for RIS-empowered narrowband communications in indoor and outdoor environments and includes many physical characteristics, such as LOS probability, shadowing effects, and shared clusters. In addition, the model incorporates realistic gains and array responses for RIS elements in addition to the existing channel models. Based on this model, an open-source channel simulator MATLAB package has been finalized, which can be used in channel modeling of RIS-based systems with tunable operating frequency, terminal locations, number of RIS elements, and environments.
%\cite{zegrar_Reconfigurable_2020} MIMO system with RIS (no direct path between transmitter and receiver). The channel is modeled as a keyhole MIMO system.

In \cite{bjornson_Rayleigh_2021}, the authors proved that channel fading in RIS-empowered communications will always be spatially correlated \cite{WeibullAWPL_2007,NakagamiTWC_2009,NakagamiTVT_2010,GammaIET_2010}, thus discouraging the common adoption of the independent and identically distributed Rayleigh fading model. It was shown that the asymptotic SNR limit is equal, but the convergence rate and rank of the spatial correlation matrices are different. A channel model for isotropic scattering was presented, whose properties were characterized, including the rank and channel hardening. The derived channel properties were also applied for massive MIMO arrays with
the same form factor, called holographic MIMO \cite{husha_LIS1,huang2020holographic}. Kronecker-product covariance matrices for the incoming and outgoing channels to/from RISs were recently proposed in \cite{Moustakas_RIS}, which relate the field's mean
direction of arrival/departure and its angle spread with the spatial correlation in planar arrays. The presented model enables the investigation of the role of the channel conditions near RISs in wireless communications empowered by multiple RISs. Based on this model, an asymptotic closed-form expression was derived for the mutual information of a multi-antenna transmitter-receiver pair in the presence of multiple RISs, in the large-antenna
limit, using the random matrix and replica theories. Under mild assumptions, asymptotic expressions for the eigenvalues and the eigenvectors of the channel covariance matrices were derived. It was found that, when the channel close to an RIS is correlated (e.g., for
instance due to small angle spread), the communication link benefits significantly from the RIS optimization, resulting in gains that are surprisingly higher than the nearly uncorrelated case. Furthermore, when the desired reflection from the RIS departs significantly from geometrical optics, the surface can be optimized to provide robust communication links. Building on the properties of the eigenvectors of the covariance matrices, the authors of \cite{Moustakas_RIS} were able to find the optimal response of the RISs in closed form, bypassing the need for brute-force optimization.

In \cite{pizzo_Spatially_2020}, the authors consider three-dimensional small-scale fading in the far-field and assume that it can be modeled as a zero-mean spatially-stationary and correlated Gaussian scalar random field satisfying the Helmholtz equation in the frequency domain. This model yields a physically meaningful spatial correlation function, whose power spectral density (in the spatial-frequency or wavenumber domain) is impulsive with support on the surface of a sphere of radius $\kappa = 2\pi/\lambda$ (with $\lambda$ being the wavelength), and can be uniquely described by a spectral factor that specifies directionality and physically characterizes the propagation environment in its most general form. This structure of the spatial correlation function leads to the two-dimensional Fourier plane-wave spectral representation for the field, which is given by a superposition of a continuum of plane-waves having zero-mean statistically-independent Gaussian-distributed random amplitudes. It is shown that the small-scale fading has a singularly-integrable band-limited spectrum in the wavenumber domain that is defined on a disk of radius $\kappa$. This is a direct result of the Helmholtz equation, which acts as a two-dimensional linear space-invariant physical filter that projects the number of observable field configurations to a lower-dimensional space. The band-limited nature of the two-dimensional Fourier plane-wave spectral representation is exploited to statistically characterize the small-scale fading over holographic MIMO arrays of compact size. Recently, the authors in \cite{Holo_Modeling_2021} extended the results of \cite{pizzo_Spatially_2020} to downlink multi-user communications, by modeling the communication channel in the wavenumber domain using the Fourier plane wave representation. Based on this channel model, maximum-ratio transmission and zero-forcing pre-coding schemes are devised capitalizing on the sampled channel variance, which depends on the number and spacing of the patch antennas in the holographic MIMO system. Analysis of spectral efficiency is also presented.

\begin{figure}[!t]
	\begin{center}
		\scalebox{0.234}{\includegraphics{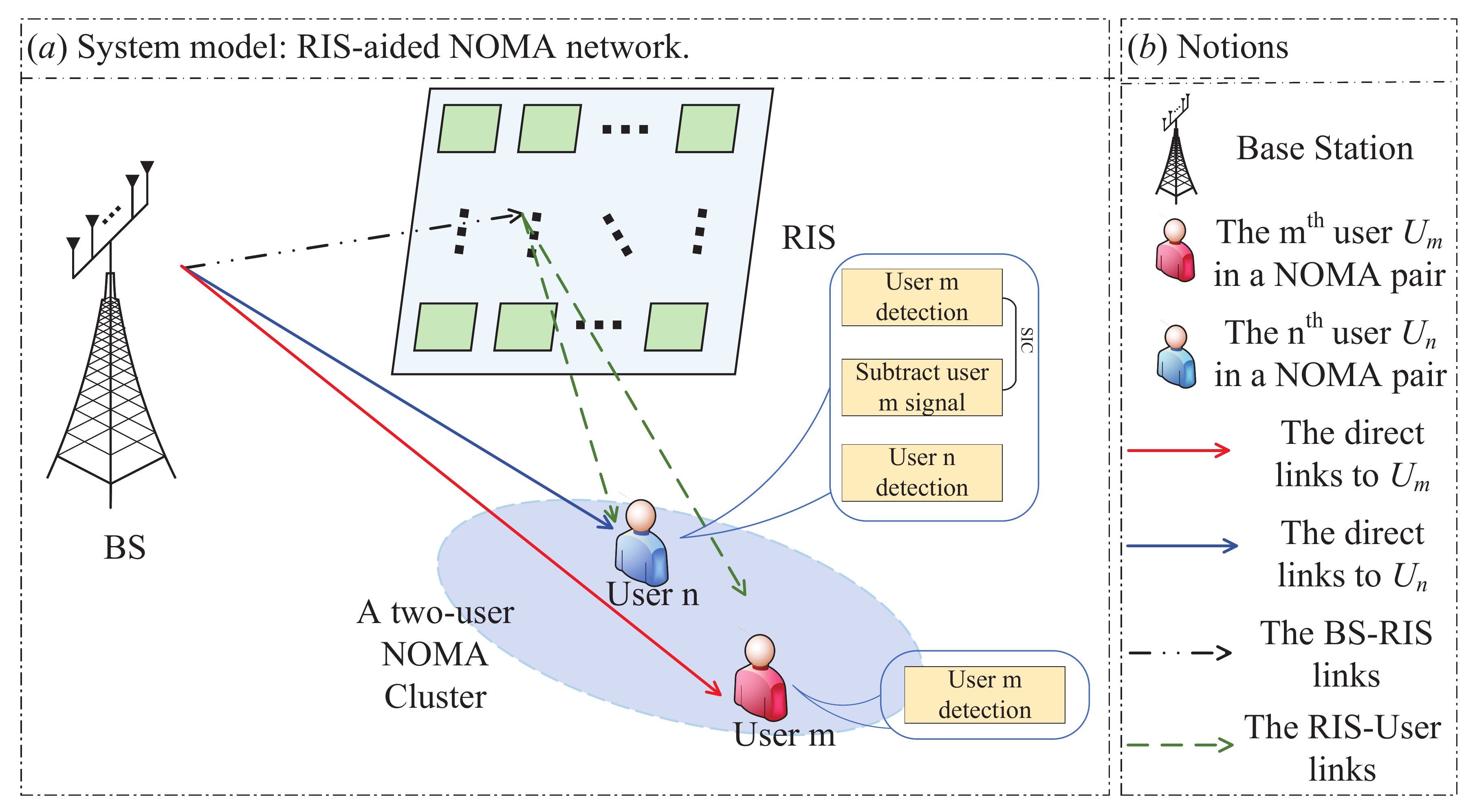}}
		\caption{Illustration of a two-user RIS-empowered NOMA system.}
		\label{Fig_RISNOMA}
	\end{center}
\end{figure}

%\subsection{Channel Distribution Analysis for RIS-Assisted NOMA}

%(RIS and NOMA introduction)
The channel characteristics in multi-user RIS-empowered NOMA communication systems has lately attracted significant research efforts \cite{Yuanwei_journal}. For the easy understanding of the readers, a simplified NOMA system with two users is illustrated in Fig.~\ref{Fig_RISNOMA}. 
%Sharing the same spectrum resources, the signals are transmitted to the two users by the direct links from the BS and the reflecting links assisted by the RISs. With the aid of successive interference cancellation (SIC), the signals can be divided to the targeted users. With the aid of the programmable phase shifts to achieve integrated signals, the advantages of RISs fill some of the emergencies of NOMA. 
%Firstly, RISs are able to enhance the performance of their aided users existing in the NOMA networks, especially for cell-edged users in downlink NOMA networks \cite{chao_NOMA_1}. Moreover, for multi-cell NOMA networks, RISs are able to reduce the interference from other cells by using the destructive interference of the reflected signals of RIS elements. From NOMA to RISs, NOMA provides efficient multiple access schemes, which enhance the spectrum efficiency with high sum rates for the RIS-aided networks. 
%Motivated by the benefits above, RIS-aided NOMA performs as a promising technology, which absorbs the investigation of RIS-aided NOMA networks on channel distribution analysis. 
In \cite{Zhiguo_NOMA_RIS}, a basic channel model was presented, where the RIS was deployed in order to ensure that additional cell-edge users can be served via spatial division multiple access beams, by aligning the cell-edge users' effective channel vectors with the predetermined spatial directions. In \cite{xu_novel}, the authors presented a model where the signals reflected through the scatters and the RIS elements are jointly studied as multipath components of the overall received envelope. Approximate distributions of the RIS-empowered channels were derived for both continuous and discrete phase shift scenarios. The authors in \cite{tianwei_NOMA_1} derived the best- and worst-case of channel statistics for characterizing the effective channel gains for RIS-aided NOMA networks, which can be used as upper and lower bounds. Additionally, \cite{ZeyuSun} investigated channel models for both ideal and non-ideal RISs. For further evaluation, \cite{yanyu_NOMA_1} estimated the diversity gains of the RIS-aided channels. Focusing on RIS channel modeling and optimization algorithms, the authors in \cite{RISNOMA3} considered RISs to deal with multiple blocked users. In \cite{RISNOMA2} and \cite{RISNOMA4}, the authors proposed a NOMA solution with RIS partitioning to enhance the spectrum efficiency.

Approximate distributions for the channel statistics in multi-cell RIS-empowered NOMA networks have been recently presented. In \cite{chao_NOMA_1}, the authors considered RISs as linear materials and introduced a path loss model for the RIS reflected channels. In addition, based on a Poisson cluster process model, the reflection angle distributions were evaluated and further used to study spatial effects. Under Rayleigh fading channels, the curve fitting channel model for RISs has been investigated by \cite{chao_NOMA_2} to mimic the integrated RIS channels as a Gamma distribution. The authors in \cite{ziyi_NOMA_2} exploit the central limit theorem to evaluate the channel models and expand the channel model approach to other distributions, such as Nakagami-$m$ channels, Rician channels, and Weibull channels with implications for their multivariate versions \cite{WeibullAWPL_2007,NakagamiTWC_2009,NakagamiTVT_2010,GammaIET_2010}.

%%%

\section{Channel Estimation in RIS-Empowered Systems}\label{sec:Channel_Estimation}
%A favorable characteristic of RIS is that it is a passive and nearly continuous surface that can realize reconstruction of EM waves. This feature gives RIS bright development prospects, especially in expanding coverage, overcoming coverage holes, enabling environmental perception and positioning, and supporting green communications.

It is indispensable to obtain accurate channel state information (CSI) to achieve intelligent control of the electromagnetic environment. The channel estimation (CE) for RIS involves the estimation of multiple channels simultaneously: the direct channel between the base station (BS) and each user equipment (UE), the channel between the RIS and BS, and the channel between the RIS and each UE \cite{qurratintelligent,9133156}. Moreover, RIS channel estimation faces several challenges, including the significantly increased channel estimation dimension, cascaded channel measurement results, and the near-field channel estimation problem. This task becomes more complex when the deployed RISs are equipped with massive elements with non-linear hardware impairments \cite{huang2020holographic,9324910}. The unavoidable hardware error decreases the accuracy of channel estimation, and the beam squint phenomenon occurs in wideband channels. Many literature on channel estimation has been published to solve the aforementioned problems, and RIS channel estimation schemes have been designed from different perspectives.  

In different application scenarios, the channel characteristics of the BS-RIS and RIS-UE channels can be different. For example, in the scenario where the RIS is deployed on the facade, the BS-RIS channel is semi-static with the coherence time being extended and main energy is concentrated on the LOS path. In this case, channel estimation can be done in a longer periodicity. Conversely, due to the mobility of the UE, the coherence time of the RIS-UE channel is significantly shorter, and the channel estimation needs to be done in shorter periodicity than that of the BS-RIS channel. 
%In this scenario, proper channel estimation scheduling and feedback using the slow-changing characteristic of the BS-RIS channel need to be designed. 
In another possible scenario where RIS is deployed on glasses of high speed trains, the BS-RIS channel is dynamic while the RIS-UE channel can be regarded as semi-static. How to perform high-dimensional BS-RIS fast-changing channel estimation becomes the major challenge. Under scenarios where the RIS is deployed on mobile objects such as buses or unmanned aerial vehicles (UAV), the coherence time of both the BS-RIS channel and the RIS-UE channel is relatively short, where rapid estimation is needed for both channels.

There are many literature investigating the framework design of CE for RIS-empowered wireless systems. In \cite{8879620}, a general framework for cascaded channel estimation in RIS-empowered MIMO systems is introduced by leveraging combined bilinear sparse matrix factorization and matrix completion. A control protocol enabling linear square estimation for the involved channel vectors that activates a part of RIS elements in the MISO case is presented in \cite{mishra2019channel}. In \cite{hu2021two}, a two-timescale channel estimation schedule is introduced which utilizes the semi-static characteristic of the BS-RIS channel. It estimates the high-dimensional BS-RIS channel in a relatively long period and performs multiple low-dimensional RIS-UE CEs under one BS-RIS channel estimation. In \cite{guan2021anchor}, two anchor nodes are deployed near the RIS to estimate the BS-RIS-UE channel and two cascaded channel estimation frameworks are presented accordingly. The first approach takes advantage of the fact that all users share the BS-RIS channel. Within the coherence time, a part of the BS-RIS channel is estimated based on anchor-assisted training and feedback, and then the cascaded channel is estimated using user pilot signals more frequently. The second approach estimates the relevant channels of the anchor point A1 and A2 first, namely the BS-RIS-A1 and A1-RIS-A2 channels, then estimates the A2-RIS-UE channel to restore the cascaded BS-RIS-UE channel. In \cite{8937491}, a transmission protocol for RIS CE and configuration optimization is proposed for RIS-empowered OFDM systems. However, the presented RIS reflection pattern is implemented by ON/OFF switching, which can be costly in practice, and the estimation accuracy suffers since only a portion of RIS elements are activated during the CE period. Recently, a novel transmission protocol with quasi-periodic training symbols is proposed in \cite{LZAWFM2020} to perform channel estimation and passive beamforming simultaneously. This adaptive protocol adopts RIS elements grouping and the time interval of adjacent training symbols can be adjusted to balance data transmission delay and effective transmission rate for supporting delay-sensitive transmissions.

The introduction of RIS makes the channel dimension increase sharply, thus it is urgent to explore a short-pilot and low-overhead channel estimation and feedback scheme, which has become one of the major issues that must be considered in the design of the actual RIS channel estimation algorithm. RIS element grouping is a possible way to reduce the pilot length. In \cite{yang2020intelligent}, a novel RIS elements grouping method is proposed as each group consists of a set of adjacent RIS elements that share a common reflection coefficient. The training overhead is greatly reduced as only the combined channel of each group needs to be estimated. Similarly, a novel hierarchical training reflection design is proposed in \cite{you2020channel} to progressively estimate the RIS-empowered channels over multiple blocks by exploiting RIS-elements grouping and partition. Moreover, the pilot length can be reduced if only the aggregated channel from the BS to the users, instead of every individual channel coefficient, is estimated \cite{van2021reconfigurable}. In the two-timescale scheme \cite{zhi2021two}, the BS beamforming is designed based on the instantaneous aggregated CSI and the length of the pilot sequences only needs to be larger than the number of users instead of the number of RIS elements. 

The design of the RIS reflection pattern affects the channel estimation performance as it controls the reflection signal. The authors in \cite{jensen2020optimal} claim that they design an optimal CE scheme, where the RIS elements follow an optimal series of activation patterns. In \cite{9081935}, the authors design the joint optimal training sequence and reflection pattern to minimize the channel estimation MSE for the RIS-empowered wireless system by recasting the corresponding problem into the convex semi-definite programming problem. However, the authors simply estimate cascaded channels instead of separate channels.

The feature utilization can be of great help to simplify the estimation procedure. The single-structured sparsity of RIS-UE channels in the hybrid space and angle domain is analyzed in \cite{shen2021dimension}. It feeds back user-independent CSI and user-specific CSI through dimension reduction method and designs a dynamic codebook through angle information to realize BS-RIS-UE cascaded channel estimation. On this basis, \cite{wei2021channel} analyzes the double-structured sparsity of the angle domain cascaded channel, using the common BS-RIS channel and the sharing environmental scatters. It reveals that the angular cascaded channels related to different users have identical non-zero rows and partially identical non-zero columns, and the channel estimation is realized through the DS-OMP algorithm. In addition, to solve the high-dimensional CE problem and avoid huge pilot overhead, \cite{kundu2021large} studies the achievable rate of the RIS-empowered SISO communication and selects the optimal number of RIS elements according to the statistical CSI to balance the power gain and the channel estimation overhead. It helps maximize the upper limit of the average achievable rate and realize low-complexity and low-rate channel adaptation.

If all RIS elements are connected to the baseband through a digital architecture, the excessive hardware complexity and power loss will affect the power cost-benefit of RIS seriously. Some research adopt the hybrid RIS hardware structure that requires a small number of active elements with sensing and signal processing ability. In \cite{Taha_2021}, a new RIS architecture based on sparse channel sensors is studied where a small number of active components are connected to the baseband. RIS channel estimation and reconstruction with significantly reduced pilot overhead are performed through compressed sensing, and deep learning tool is adopted to design the reflection matrix to achieve a rate near the upper limit. The CSI acquisition under the hardware design of the uniformly distributed active sensor is analyzed in \cite{jin2021channel}, and the rank-deficient structure of the RIS channel is utilized to design the residual neural network. The single-scale enhanced deep residual network reduces the RIS hardware complexity significantly. The multi-scale enhanced deep residual network uses the correlation between multi-scale models to ensure that the proportion of active elements is adjusted to adapt to different scenarios and achieve multi-scale super-resolution. For comparison, \cite{schroeder2021passive} studies the channel estimation performance of the fully passive RIS architecture and the hybrid RIS architecture. For fully passive RIS, the AOD of BS and the AOA of UE are estimated in the first stage, and the other channel parameters are estimated subsequently. For hybrid RIS, alternate uplink and downlink training are assumed to estimate the segmented channels.

RIS has numerous applications in the millimeter-wave frequency band, while a prominent feature of the high-frequency scenario is that the transmission signal faces severer attenuation. The path with multiple reflections is usually ignored, and only the direct path and a small amount of single-reflection path can reach the receiver. High-frequency channel usually has sparse characteristic, being significantly helpful for simplifying the RIS CE. In \cite{ardah2021trice}, a two-stage channel estimation for parameter decoupling is considered. The AOD of BS-RIS and the DOA of the RIS-UE channel are estimated in the first stage. The parameters estimated in the first stage are used to estimate the azimuth, elevation, and path gain of the RIS cascaded channel in the second stage. Low-overhead CE is achieved through ESPRIT or compressive sensing algorithm. The inherent sparse quasi-static BS-RIS channel assumption is adopted in \cite{9133156}, and the long-term statistical information of the slow-varying channel components are utilized to carry out jointly BS-RIS and RIS-UE channel calibration and estimation. The cascaded channel can be inferred through Bayesian posterior estimation, and the approximate message passing (AMP) algorithm is used to reduce the complexity of estimation approximation. Moreover, \cite{mirza2021channel} considers the ill-conditioned low-rank channel where RIS is deployed near the UE, transforming the RIS CE problem into a dictionary learning problem. The reliable and robust bilinear adaptive vector AMP algorithm is used to estimate the BS-RIS-UE channel, and the sparsity of the RIS phase shift matrix is utilized in the training phase to eliminate the restored channel permutation ambiguity. A multi-objective optimization problem based on the joint received signal nuclear norm and the channel gain-norm objective is proposed in \cite{liu2021admm}, and the cascaded channel estimation is transformed into a sparse matrix recovery problem. The channel estimation method based on the alternating direction multiplier achieves high accuracy CE performance.

An important advantage of high-frequency transmission is that the available bandwidth is large. However, the spatial wideband effect may occur in millimeter-wave and THz frequency bands, affecting channel estimation accuracy and bringing about dispersion problems in beam regulation \cite{9498810}. Additionally, the large surface of RIS will further aggravate the delay between the signals received by RIS elements, which may cause different RIS elements to receive different symbols. Correspondingly, different subcarriers may point towards different directions for transmitting and cause dispersion problems. It is proved in \cite{ma2021wideband} that the cross-correlation function between the spatial steering vector and the cascaded channel has two peaks in the angle domain, that is, the frequency-dependent actual angle and the frequency-independent error angle. The existence of the error angle affects estimation performance severely, and a two-stage orthogonal matching pursuit (TS-OMP) algorithm is designed accordingly. The influence of beam squint on transmission is discussed in \cite{liu2021cascaded}, transforming the wideband CE into the recovery problem of angle, delay, gain, and other parameters. The Newton OMP algorithm is proposed to detect the channel parameters, and the Cramer-Rao lower bound (CRLB) is derived to provide strong support for performance evaluation. In \cite{chen2021beam}, the problem that different incident angles bring different path phases due to beam squint while the RIS phase shift matrix has the same value for all frequencies is analyzed. The phase shift matrix is designed based on the SVD decomposition of the average channel covariance matrix to reduce the negative effect of beam squint effect on system performance.

Another direction to solve the high-dimensional CE problem is channel splitting. The high-dimensional channel can be equally split into multiple low-dimensional channels, greatly reducing channel estimation complexity. An ON-OFF-based MMSE channel estimation method is designed in \cite{al2020intelligent}, splitting the RIS into multiple elements and setting them as ON state at different times. CE is performed for each RIS element separately, achieving low complexity channel estimation. It is assumed in \cite{jensen2020optimal} that there is no prior knowledge about the channel and linear square estimation is used to estimate channel through each RIS element (or group). It aims to minimize CRLB under the constraints of attenuation and phase quantization models, and the optimal design is guided by the minimum variance unbiased estimation criteria. Also, it shows that the estimation performance depends on the activation mode of RIS. Actually, the ON/OFF method \cite{al2020intelligent,mishra2019channel,qurratintelligent} generally yields sub-optimal CE results. %The best RIS activation mode follows the discrete Fourier transform matrix to achieve the estimated variance decrease by orders of magnitude. 
In \cite{zegrar2021reconfigurable}, the channel passing through each RIS element is modeled as a keyhole channel, and the rank-one feature is considered to simplify the estimation. This paper designs a channel splitting expansion scheme and uses eigenvalue decomposition to perform cascaded channel estimation for RIS-empowered communications, greatly reducing the estimation time overhead. Similarly, a sub-channel estimation scheme is designed \cite{zhou2020joint} based on multiple rounds of pilot training. It appropriately configures the RIS phase shift matrix as hadamard matrix and DFT matrix and uses sub-channel CSI to design joint RIS-transmitter precoding. 

In \cite{de2021channel}, a tensor modeling approach is considered to transform the CE problem of RIS-empowered communications to a 3D tensor through the parallel factor (PARAFAC) model and perform channel estimation using cascaded channel decoupling. It derives a closed-form solution based on the Khatri-Rao decomposition of the cascaded channel. An iterative alternate estimation scheme promotes the tensor signal model to the BS and multiple UEs scenarios. Similarly, inspired the promising results of the PARAFAC decomposition \cite{harshman1994parafac} in CE for relay systems, two iterative estimation algorithms through PARAFAC of cascaded channels are proposed in \cite{wei2021channel1}. Alternate least square (ALS) and vector approximation message passing (VAMP) algorithms are designed to reconstruct the unknown segmented channels. This paper gives feasibility conditions and computational complexity evaluation, showing promising CE performance gains. In \cite{tensor_channel_tracking_2022}, the adaptive PARAFAC decomposition scheme is adopted to get the initial estimate and track the time-varying channel for the uplink of RIS-enabled MU MIMO systems. Moreover, the Generalized Approximate Message Passing (GAMP) is utilized to recover the RIS-UE channels with reduced the pilot overhead.

The mobility of the UE brings challenges to channel estimation and it is inevitable to study the time-varying channel in a mobile scenario. If the mobility cannot be well handled, it becomes challenging to avoid channel estimation deviations, leading to channel matching regardless of link adaptation or beamforming and resulting in a gap between the actual achievable capacity and the ideal channel capacity. In \cite{mao2021channel}, the cascaded channel is represented as a mobile state-space model. A Kalman filter-based scheme is designed to track the time-varying channel and improve estimation accuracy using the channel time correlation and prior information. The time-varying cascaded channel estimation problem is exploited in \cite{xu2021deep} by designing a deep learning channel extrapolation scheme. It decomposes the neural network into the time-domain section and the antenna-domain section. In the time domain, the dynamic channel is accurately described by combining the neural network and the cyclic neural network. In the antenna domain, the neural network is modified, the structure of the feedforward neural network is changed, and the enhanced feedforward neural network is designed to enhance the network performance. Moreover, the doppler effect is considered in \cite{sun2020channel}, and a wideband CE scheme is designed based on doppler shift adjustment for multipath and single path propagations. A quasi-static channel estimation mechanism is proposed for multipath scenarios, adjusting doppler distortion through the joint RIS phase shift matrix design and time-frequency domain conversion.

The fixed position information of BS and RIS helps design a low-complexity channel estimation method by obtaining critical information such as the AOA and AOD. The introduction of position information can also enhance other channel estimation algorithms, providing additional information, improving channel estimation accuracy, and reducing channel estimation complexity. In \cite{he2021leveraging}, the position information of the external positioning system is used to design the directional beam. The atomic norm optimization method extracts the channel angle and other parameters to achieve the estimation performance with a predetermined beam codebook. It effectively accelerates beam alignment and the channel parameter estimation process. The RIS-empowered SISO multi-carrier system is considered in \cite{keykhosravi2021siso}, and the large-scale, multi-element, and low-cost characteristics of RIS are considered for joint 3D positioning. A low-dimensional parameter space search algorithm is designed to reduce the dimensionality of the four-dimensional parameter estimation problem. It performs two one-dimensional searches for times-of-arrival and a two-dimensional search for angles and achieves sub-meter positioning and synchronization accuracy. In \cite{liu2021reconfigurable}, the influence of RIS on the positioning is analyzed from the perspective of how to design RIS phase shift matrix to improve positioning accuracy. This paper verifies the positioning channel model of the three-dimensional RIS-empowered communication system and derives the CRLB. It designs RIS passive beamforming to minimize the CRLB, solved the problem through alternate optimization and gradient descent methods, and realizes centimeter-level positioning targets.

\begin{table*}
\centering
	\caption{Complexity Analysis for RIS-Empowered MISO Communication Systems.}
	\label{Complexity}
	\begin{tabular}{|p{20pt}|p{200pt}|p{200pt}|}
		\hline\hline
		Source     & Algorithm & Complexity \\
        \hline
        \cite{mishra2019channel}   & Least Squares  &  $\mathcal{O}((N M)^2 T_p)$\\
        \hline
        \cite{de2020parafac}   &  Bilinear Alternating Least Squares (BALS) &  $\mathcal{O}(2N+4N M T_p)$\\
        \hline
		\cite{wang2020compressed}   &  Orthogonal Matching Pursuit (OMP) &  $\mathcal{O}(N M T_p)$\\
        \hline
        \cite{wei2021channel} & Double-structured Orthogonal Matching Pursuit (DS-OMP) & $\mathcal{O}(K N T) + \mathcal{O}(L_G K M T L_{k}^3)$\\           
        \hline
        \cite{jensen2020optimal} & Cramer–Rao lower bound ON/OFF & $\mathcal{O}(N^3 M^3)$\\
        \hline
        \cite{ardah2021trice} & Two-Stage RIS-Aided Channel Estimation (TRICE) - CS & $\mathcal{O}(LK_T(\bar{L}_T \bar{L}_R + L + L^2) +2L^3+L K_S \bar{L}_S^v \bar{L}_S^h)$\\ 
        \hline
        \cite{zhang2017channel} & TRICE - Beamspace ESPRIT & $\mathcal{O}(K_T^2 K_S + K_S^3 +3L^3 +L)$\\    
        \hline
        \cite{9133156} & Matrix Calibration based CE & $\mathcal{O}(I_{max}(NKT + 2NKL_G +N N_G (L_G)^2)))$\\ 
        \hline
        \cite{han2020deep} & DoA with 1-Bit Quantization & $\mathcal{O}(4BstN^2 M^2)$\\
        \hline
        \cite{liu2021admm} & Alternating Direction Method of Mmultiplier (ADMM) & $\mathcal{O}(I_{max} N M^2 T_p)$\\
        \hline
        \cite{hu2021two} & Coordinate Descent-based Channel Estimation & $\mathcal{O}(M^3+MLN^2+MLNI_{max})$\\         
        \hline
        \cite{jin2019channel} & Convolutional Blind Denoising Network (SBDNet) & $\mathcal{O}(NMK^2 st(L_d D_l^2 + L_e E_l^2))$\\       
        \hline
        \cite{jin2021channel} & Single-scale Enhanced Deep Residual (EDSR) & $\mathcal{O}(4NM K^2 (2B+\kappa^2)st D^2)$\\
        \hline
        \cite{jin2021channel} & Multiple-scale Enhanced Deep Residual (MDSR) & $\mathcal{O}(8NM K^2 (B+16)st D^2)$\\
        \hline
        \cite{wei2020parallel} & Alternating Least Squares (ALS) & $\mathcal{O}(2M^3+4M^2 P(K+N)-M P(K+N)$\\     
        \hline
        \cite{wei2021channel1} & Vector Approximate Message Passing (VAMP) & $\mathcal{O}((K+N)(5M^2-M))$\\
        \hline
        \cite{sun2020channel} & Multi-path CE Incorporating Doppler Shift Adjustment & $\mathcal{O}(M^2 N_O +MN_O \log N_O)$\\      
        \hline
        \cite{kundu2021channel} & Denoising Convolutional Neural Network (DnCNN) & $\mathcal{O}(N^2(M+1)^2+9NN_f(M+1)(4+(D-1)N_f))$\\
        \hline
        \cite{kundu2021channel} & Fast and Flexible Denoising Network (FFDNet) & $\mathcal{O}(N^2(M+1)^2+9NN_f(M+1)(4.5+0.5(D-2)N_f))$\\
        \hline\hline
	\end{tabular}
\end{table*}

The rapidly developed machine learning approaches have been lately presented \cite{huang2019spawc,Taha_2021,hardware2020icassp} to reduce the pilot overhead and computational complexity in RIS-empowered systems. Considering an indoor scenario, a deep neural network (DNN) is designed to unveil the mapping between the measured coordinate information at a certain user location and the configuration of RIS elements to maximize the received signal strength of a given user \cite{huang2019spawc}. In \cite{Taha_2021}, the authors consider the RIS with some active elements for partial channels sensing. They present an approach based on compressive sensing to recover the full channels from the samples. The rich-scattering environment is considered in \cite{Jointly_Learned_2021} and a deep neural network is trained as surrogate forward model to capture the stochastic dependence of such RIS-empowered wireless channels. The link between the RIS configuration and the key statistical channel parameter is captured to determine the communication rate with the help of a DNN. Recently, in \cite{hardware2020icassp}, a RIS architecture consisting of any number of passive RIS elements, a signal controller, and a single RF chain for baseband measurements is presented. This novel architecture is proposed to estimate channels at the RIS side via matrix completion for sparse channels. 

However, these machine learning based methods \cite{huang2019spawc,Taha_2021,hardware2020icassp} either require extensive offline training or is built on the assumption that RISs have some active elements to realize analog or hybrid reception. Intuitively, architectures based on the latter assumption would mean an increase in the RIS hardware complexity and power consumption. In \cite{kundu2021channel}, the approximate optimal MMSE channel estimation scheme is studied, and the best linear LMMSE estimation based on the majorization-minimization algorithm is analyzed. A data-driven nonlinear solution based on deep learning is proposed, achieving denoising and approximating the optimal MMSE through convolutional neural networks. In \cite{shtaiwi2021ris}, the number of active users is decreased to reduce pilot overhead, and a two-stage scheme for overall CE is designed. In the first stage, the AOA at BS is obtained through energy detection, the effective path gain is obtained through the LS method, and the AOA of the RIS-UE channel is derived correspondingly. In the second stage, partial CSI obtained in the first stage is used to handle the spatio-temporal correlation between adjacent users. The mapping of the channels of active and inactive users can be realized. The channel extrapolation method of activating part of RIS elements is studied in \cite{gao2021deep}. The proposed three-stage CSI acquisition of BS-UE comprises direct channel estimation, partial RIS cascaded channel estimation, and overall cascaded channel prediction through deep neural networks. The cumulative estimation error of DNN training in the prediction stage can be overcome effectively through the BS antenna mapping relationship. Moreover, in \cite{DeepRIS_2022}, a ML-based receiver is designed by treating the RIS reflection pattern as hyperparameter and a Bayesian ML framework is optimized to jointly tune the RIS and the multi-antenna receiver.

The analysis of RIS channel estimation and feedback mainly focus on the design of the ideal RIS hypothesis, but there are inevitable errors in practical applications. For example, the ideal RIS hypothesis assumes that each RIS element has a constant amplitude, variable phase, giving the same response to different frequencies. The discrete phase shift model and the relationship between the response under actual RIS hardware are analyzed in \cite{yang2021channel}. LS channel estimation scheme based on non-ideal RIS hardware is designed, and an alternate iterative algorithm is proposed to obtain the RIS time-varying phase shift matrix to minimize the NMSE. In \cite{liu2021channel}, RIS hardware impairments are considered, and a linear LMMSE estimator is designed. The relationship between channel estimation performance and impairment level, the number of reflective elements, and pilot power is given. It proves that the hardware impairment of the transceiver under a high SNR limits the CE performance. The non-ideal CSI with the random phase noise is studied in \cite{qurratintelligent}, and only large-scale statistics are utilized for BS precoding and RIS reflecting matrix design. A closed-value expression of the achievable spectrum efficiency of the uplink transmission is derived and optimized using the maximum ratio combining. The random blocking problem of RIS due to environmental influences is considered in \cite{li2020joint}, and the array diagnosis in the case of passive RIS is discussed. The non-ideal channel estimation is modeled as joint antenna diagnosis, and channel estimation is solved by constructing a two-timescale non-convex optimization problem. A batch processing algorithm is selected to identify induced attenuation and phase shift caused by blockage to achieve array blocking coefficient calculation and effective sparseness channel parameter estimation.

Evaluating the impact of CSI error on beamforming performance is the critical point of RIS-empowered communication system design. Robust beamforming based on non-ideal cascaded channel is investigated in \cite{zhou2020framework}. For the bounded CSI error model, the BS precoding matrix and the RIS phase shift matrix are jointly designed to minimize the total transmission power under the constant modulus constraint and the worst-case QoS constraint. For the statistical CSI error model, modulus constraint and rate interruption probability constraint are considered to achieve better performance with minimum transmit power, low convergence speed and complexity. In \cite{jung2020performance}, the channel hardening effect is analyzed, assuming that the interference channel is the spatially correlated Rician fading. The UE uplink data rate and the RIS system performance limit are derived. It concludes that the hardware impairments and noise from estimation errors and the non-line of sight (NLOS) path could be ignored with massive RIS elements. The statistical CSI of the BS is utilized in \cite{zhang2021large} to design the best transmission covariance matrix and RIS phase shift matrix. The achievable data rate is optimized through iterative water-filling optimization and the RIS phase shift matrix is optimized by the projected gradient ascent method.

In Table \ref{Complexity}, we analyze the complexity of different types of channel estimation algorithms, hoping to provide a basis for channel estimation algorithm selection in actual system design. As can be observed, the AI-based channel estimation schemes can exploit the massive channel measurement data and achieve good channel estimation performance with acceptable complexity. With the continuous development of AI technology, AI-based channel estimation can have greater development prospects in the future. Moreover, we found that in-depth analysis of the special properties of RIS-empowered communication can effectively reduce the complexity of channel estimation, which is also a brighting way for future channel estimation design.

\section{RIS Standardization:\\ Current Status and Road Ahead}\label{sec:Standardization}
In this section, some highlights on the early standardization efforts on RIS are summarized and presented. \cite{9538911} provides some details related to standardization activities as well.

As extensive research efforts are invested on RISs globally, the standardization work has also taken off, mainly on a regional-level.
In China, one of the largest single market for wireless devices, RIS-related standardization work has already been initiated in two different organizations, namely, the FuTURE Mobile Communication Forum and the China Communications Standards Association (CCSA). 
In the FuTURE forum, a working group (WG) to study approaches to integrate RISs into next generation wireless networks was established in December 2020 \cite{futureforum}. This WG studies the use cases and key technologies to support RISs towards commercialization, and a white paper is planned to summarize the technical trends for further promoting RIS-assisted wireless systems.
Compared to the FuTURE forum, CCSA is a more formal standards developing organization (SDO) which can produce normative specifications instead of on top of informative reports.  
During its 55th meeting, the CCSA technical committee 5 - working group 6 approved a proposal to establish a study item (SI) on RIS \cite{ccsa_2020}. The SI aims to conclude in June 2022 with a technical report as the major deliverable. The scope of this SI includes channel modeling, channel estimation and feedback, beamforming with RISs, AI and RISs, as well as network protocols for RIS-assisted networks.
In June 2021, a new industry specification group (ISG) on RIS was approved by the European Telecommunications Standards Institute (ETSI). During the second plenary meeting of ISG RIS, three new work items were approved, which focus on channel and system modeling, use cases and deployment scenarios, and impact to current standards, respectively. All work item descriptions and related contributions can be found on the ETSI portal \cite{ETSI_newWI}. The draft group reports of the 3 WIs are targeted to be released in the middle of 2022 and stable report shall be made available by the end of 2022. The ISG will focus on studies and generating informative reports till the year of 2023, after which normative specifications may be considered.

The regional efforts to standardize RISs have some echos in global SDOs. During the ITU-R WP 5D meeting in October 2020, countries submitted their candidate proposals for next generation wireless technologies, which will be considered for inclusion into the IMT Future Technology Trends report. According to the draft report \cite{ITU_report}, RISs are depicted as critical components for the physical layer of next generation networks. In 3GPP, there was a first proposal submitted to 3GPP during the March 2021 meeting \cite{zte_proposal} by ZTE Corporation, after which more and more companies are also showing interests in supporting RIS to become a key component of 5G-Advanced networks and beyond, especially the network operators.

\begin{figure}[!t]
\begin{center}
		\scalebox{0.33}{\includegraphics{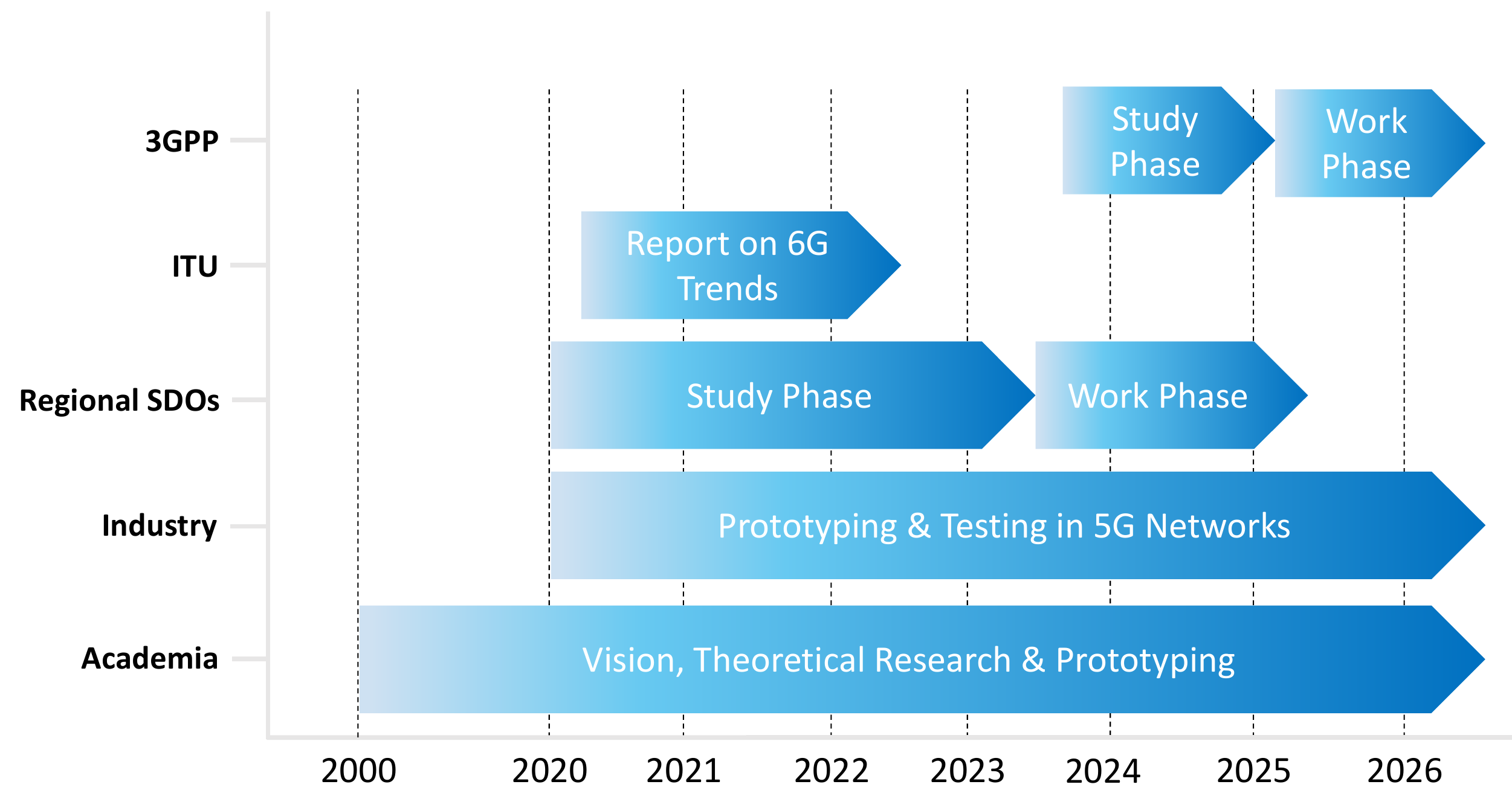}}
		\caption{Current status and forecasted timeline for the standardization of RISs.}
		\label{timeline}
	\end{center}
\end{figure}

Under the current status quo, the following future progress is predicted as depicted in Fig.~\ref{timeline}. In regional SDOs, the standardization of RISs is under way already as described in the previous part of this section. For future standard activities, it is unlikely for RIS to be established as a dedicated focus group or a WI in the ITU since the ITU mainly focuses on regulatory, spectrum and business aspects. However, RIS related channel modeling is possible to be studied in the ITU. As far as the 3GPP is concerned, the current focus is starting the second stage of 5G by Release 18, and there is no formal plan for 6G yet. According to past experience, each generation of wireless communication standards usually lasts for a decade, indicating that it is reasonable to expect that possible discussions on requirements for 6G systems will begin after 2027. Based on the current standardization progress on 5G, we estimate the first 6G release will be between Release 20 to Release 22. With the COVID-19 pandemic and its profound impact to the global economy and society, the future schedule of 3GPP may be delayed. To standardize RIS as a whole new technology, some preparation work needs to be done and one of the most important work is channel modeling \cite{liu2021path}. Due to the novel structure of metamaterials, new channel models are required to model the channel between the antenna of the base station or the UE and reflective elements or regions on RIS.

In 3GPP, there are two possible ways to standardize RISs. One is to first initiate an SI on use cases, deployment scenarios and channel models in Release 19 and then to initiate a WI after the completion of the corresponding SI. This would allow RISs to be deployed as a component of 5G networks, which will help maturing this technology and manufacturing processes. In this case, RISs will be part of 5G standards and, therefore, will be automatically inherited to 6G standards. Another possible approach is to standardize RISs as part of 6G standards together with other new features for 6G. This implies that the deployment of RISs will be delayed to around 2030s and is subject to uncertainties. Since the RIS technology can be seen as generic and band-agnostic, standardizing it as early as possible would enable opportunities for RISs to be combined with other technologies, thus contributing to greener, safer and more reliable wireless networks. On the other hand, there are still some challenging problems to be tackled, which requires a certain amount of time and efforts. Considering the current research status and the maturity of this technology, the year to start studying RISs in 3GPP is likely to be 2023. 

The current cellular and local area networks are designed to maximize the spectrum efficiency and energy efficiency, resulting in advanced and complicated physical layer designs and higher layer mechanisms and signalling. To study and standardize RIS, it is best to start with simple scenarios and practical models. For instance, considering only fixed RIS panels would avoid mobility issues and help to stabilize the network topology. What's more, although carrier aggregation is widely used in different generations of cellular networks, single carrier operation can be prioritized and studied first. Some key aspects of standardization also relates to specific frequency bands on which RIS will be deployed, for example the channel modeling and estimation, and it is more efficient to consider lower frequency bands as a starting point.

\section{Conclusions}\label{sec:Conclusions}
Due to the increased potential of RISs for 6G wireless communication networks, as witnessed by the various recent proof of concepts ranging from passive reflectarrays and dynamic metasurface antennas, there has been lately surprisingly increasing research and development activities on the RIS topic from both academia and industry working in antenna design, metamaterials, electromagnetics, signal processing, and wireless communications. The incorporation of RISs in wireless networks has been recently advocated as a revolutionary means to transform any wireless signal propagation environment to a dynamically programmable one for various networking objectives, such as coverage extension, environmental perception and positioning, and spatiotemporal focusing. In this paper, we provided an overview of the latest advances in RIS hardware architectures and their operational considerations, 
as well as the most recent developments in the modeling of RISs and RIS-empowered wireless signal propagation. We also presented the up-to-date channel estimation approaches for RIS-empowered communications systems, which constitute a prerequisite step for the optimized incorporation of RISs in future wireless networks. We discussed the relevance of the RIS technology in the latest standards, and highlighted the current and future standardization activities for RISs and RIS-empowered wireless networking.

\ifCLASSOPTIONcaptionsoff
  \newpage
\fi

\bibliographystyle{IEEEtran}
\bibliography{Reference}

% that's all folks
\end{document}